\documentclass[preprint]{epl}

\title{Correlation of tunneling spectra with surface nano-morphology and doping in thin \chem{YBa_2Cu_3O_{7-\delta}} films}
\shorttitle{Correlation of tunneling spectra...}

\author{A. Sharoni\inst{1} \and G. Koren\inst{2}
        \and O. Millo\thanks{Corresponding author, e-mail:
        \email{milode@vms.huji.ac.il}}\inst{1}}

\institute{
  \inst{1} Racah Institute of Physics, The Hebrew University, Jerusalem 91904,
  Israel\\
  \inst{2} Department of Physics, Technion - Israel Institute
           of Technology, Haifa 32000, Israel
}

\pacs{74.50.+r}{Proximity effects, weak links, tunneling
 phenomena, and Josephson effects}
\pacs{74.72.Bk }{Y-based cuprates}
\pacs{74.80.-g} {Spatially inhomogeneous structures}

\begin{document}

\maketitle

\begin{abstract}
Tunneling spectra measured on thin epitaxial
\chem{YBa_2Cu_3O_{7-\delta}} films are found to exhibit strong
spatial variations, showing U and V-shaped gaps as well as zero
bias conductance peaks typical of a \emph{d}-wave superconductor.
A full correspondence is found between the tunneling spectra and
the surface morphology down to a level of a unit-cell step.
Splitting of the zero bias conductance peak is seen in
optimally-doped and overdoped films, but not in the underdoped
ones, suggesting that there is no transition to a state of broken
time reversal symmetry in the underdoped regime.
\end{abstract}

The \emph{d}-wave nature of superconductivity in
\chem{YBa_2Cu_3O_{7-\delta}} (YBCO) and some other high-$T_c$
superconductors is well established by now \cite{1,2}.  In
particular, various groups have reported directional tunneling and
point contact spectroscopy measurements that exhibit a dominant
$d_{x^{2}-y^{2}}$ symmetry for the order parameter of YBCO
\cite{3,4,5,6,7,8}. However, the spatial variation of the
electronic properties in the superconducting state and its
correlation with the surface morphology has not been reported for
this system. These properties are reflected in the quasi-particle
density of states (DOS), which can be measured by tunneling
spectroscopy \cite{9}.  Such studies are important for resolving
various open problems regarding the nature of superconductivity in
YBCO. For example, contrary to theoretical predictions
\cite{10,11}, zero-bias conductance peaks (ZBCP) due to surface
Andreev bound state (ABS) were commonly observed when tunneling
was nominally in the [100] direction \cite{3,4,5}.  This was
attributed, but not yet demonstrated experimentally, to faceting
and microscopic roughness that reveal the (110) face at which ABS
are expected to reside \cite{3,5,12}. Another intriguing issue is
that of the emergence of a state with broken time reversal
symmetry at the surface due to the existence of a subdominant
order parameter with $\pi/2$ phase difference relative to the
dominant $d_{x^{2}-y^{2}}$ symmetry \cite{12,13}. This phenomenon
manifests itself as a splitting of the ZBCP in zero magnetic field
at low enough temperatures.  Such a splitting was observed by some
groups \cite{4,5} but not by others \cite{6, 8, 14} and it was
shown by Deutscher that the splitting might depend on the doping
level \cite{5, 15}.  Therfore, since the doping level may
fluctuate spatially over the surface even on a sub-micrometer
scale \cite{16, 17}, the nature of the ZBCP and its splitting
should consequently exhibit strong spatial variations. In the
present study we exploit the high spatial resolution enabled by
scanning tunneling microscopy (STM) and spectroscopy for the
investigation of the above issues.

Extensive STM studies of high-$T_c$ superconductors have been
performed in recent years \cite{18}.  These include atomically
resolved investigations of the role of disorder in YBCO \cite{19},
as well as the effects of atomic-size impurities on the local
quasi-particle DOS in a single crystal of \chem{BiSr_2CaCu_2O_8}
(BSSCO) that portray the \emph{d}-wave nature of this
superconductor \cite{20, 21}. However, the effect of larger scale
topographic features, such as micro-facets and unit-cell steps, on
the local DOS was not yet treated, and this is the main focus of
our present STM study of ultra-thin YBCO films. Our tunneling
spectra exhibit pronounced spatial variations that are well
correlated with the surface morphology. We show that a single
unit-cell step on the \emph{ab}-surface is sufficient to induce a
ZBCP when tunneling along the \emph{c} axis. Moreover, splitting
of the ZBCP was found in various locations (but not everywhere) on
nominally (as prepared) overdoped and optimally-doped samples, but
was never observed in the underdoped films, consistent with ref.
\cite{15}.

\begin{figure}
\onefigure [scale=0.75]{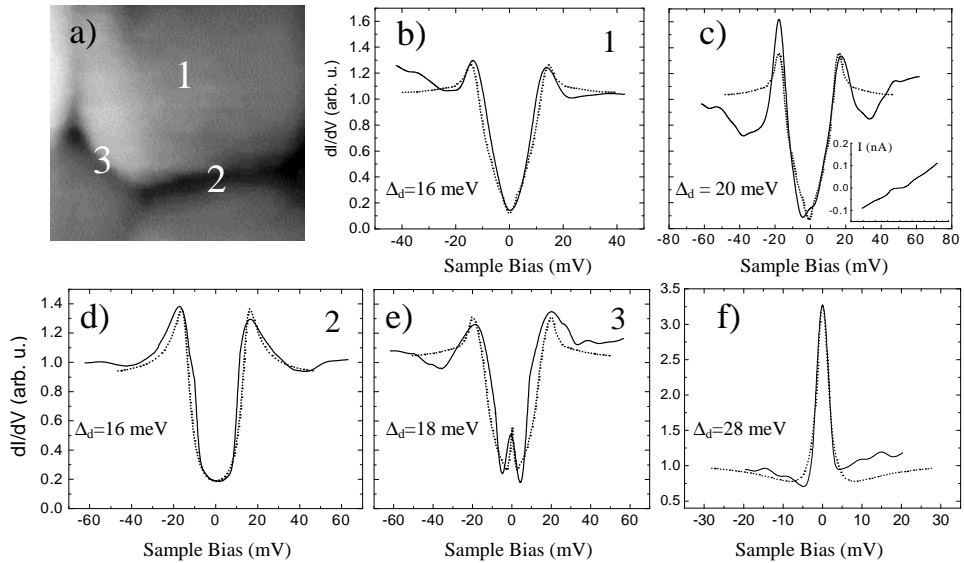} \caption{Correlation between
tunneling spectra (at $4.2 ~K$) and the surface morphology of a
thin YBCO film. (a) $50x50 ~nm^{2}$ STM topographic image showing
four crystallites and the interfaces between them. The numbers 1-3
denote the positions where spectra b, d and f were acquired. (b-f)
Tunneling spectra measured at different locations (black lines),
where (b-e) were taken on the same sample and (f) on a different
one.  The dotted curves are fits to the theory of tunneling into a
\emph{d}-wave superconductor with order parameters, $\Delta_{d}$,
denoted in the figures. The spectra exhibit various structures
that manifest \emph{d}-wave superconductivity: (b) V-shaped gap
for the curve measured on top of a crystallite (position 1),
typical for \emph{c}-axis tunneling. (c) Asymmetric V-shaped gap
with a small zero-bias structure (measured on top of another
crystallite); the I-V curve is presented in the inset. (d)
U-shaped gap for the spectrum taken at the interface position 2,
where the (100) face is revealed. (e) ZBCP inside a gap structure
for the curve measured near a corner position 3, showing
contributions of (100), (110) and (001) faces. (f) A pure ZBCP
structure taken on a (110) facet.} \label{Fig1}
\end{figure}

\begin{figure}
\onefigure [scale=0.75]{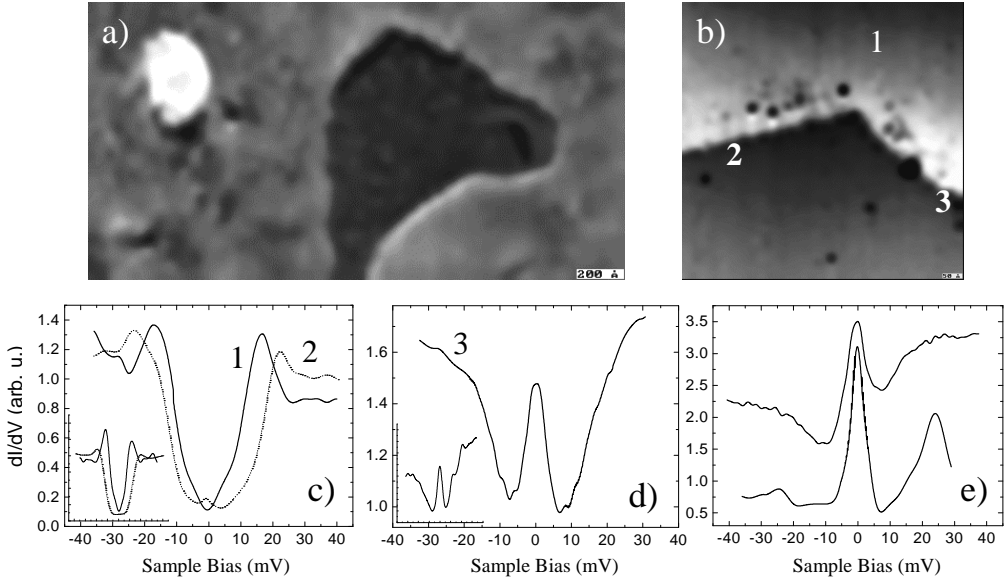} \caption{The effect of steps
of unit-cell height on the tunneling spectra measured on a
slightly overdoped film. (a and b) Topographic images of size
$120x220$ and $120x120 ~nm^{2}$, respectively, showing monocell
steps, islands and voids on the (001) surface. (c and d) Tunneling
spectra (at $4.2 ~K$) taken on different locations as denoted in
image b. Curve 1, measured far from the step, has a V-shaped gap,
manifesting \emph{c}-axis tunneling. Curves 2 and 3, measured near
the steps, exhibit ZBCPs of different magnitudes inside U and
V-shaped gaps, respectively. (e) Two highly asymmetric tunneling
spectra showing pronounced ZBCPs, taken at a different step on the
same sample. To demonstrate the spatial variation of the spectra
within this scan range we plot in the insets of (c) and (d)
additional data acquired at analogous positions as those presented
in the main frames.} \label{Fig2}
\end{figure}

Epitaxial YBCO films of $200-300$ \textrm{\AA} ~thickness were
grown on (100) \chem{SrTiO_3} wafers by laser ablation deposition,
with c-axis orientation normal to the substrate. Films of
different doping levels were obtained by the use of different
annealing conditions immediately after the deposition process.
Annealing at 0.8 atm oxygen pressure with a dwell of 1h at 450 C
yielded optimally or slightly overdoped films with $T_c \sim 90
~K$, while a cool-down under 20 mTorr oxygen flow at 900 C/h
yielded an underdoped phase of $T_c \sim 60-70 ~K$. The $T_c$ of
our films was uniform up to about $0.5 ~K$, as measured at 10
different points over the $1 ~cm^{2}$ area of the wafer. X-ray
diffraction analysis and the structure of the R(T) curves (not
shown here) confirmed the doping levels of these phases. At a
relatively low deposition rate and small laser power, the films
consisted of crystallites having well defined facets, as depicted
in fig. \ref{Fig1}a. The crystallite structure became less clear
upon increasing of the laser power, and the dominant surface
structures revealed by the STM images were voids and island of one
unit-cell height (fig. \ref{Fig2}a). The rich surface morphology
of these films enabled the observation of a wealth of spectral
structures and the study of their spatial dependence, as shown
below. The films were taken out from the growth chamber and
transferred in a dry atmosphere to a cryogenic STM within a few
hours.  The onset of the superconducting transition did not change
during this transfer process, but the transition width increased
considerably from about $1 ~K$ for freshly prepared films to
around $10 ~K$ after completing the STM measurements.  The
tunneling spectra (dI/dV vs V characteristics) were acquired
either directly by the use of conventional lock-in technique, or
by numerical differentiation of the measured I-V curves, with
similar results obtained by both methods. We have checked the
dependence of the tunneling spectra on the voltage and current
setting (\textit{i.e.}, the tip-sample distance) and found it not
to affect the measured gap and ZBCP features.  However, the
setting did influence the background, in particular the degree of
asymmetry the spectra exhibit, as noted previously by various
groups \cite{new alff}.

A typical correlation between the tunneling spectra and the local
surface morphology is depicted in fig. \ref{Fig1}, measured on an
optimally-doped film, which focuses on a nearly square shaped
crystallite about $50 ~nm$ long and $17 ~nm$ high (fig.
\ref{Fig1}a).  On top of this crystallite, V-shaped gap structures
were observed, as shown by the curve acquired at position labeled
1 in this image, presented in fig. \ref{Fig1}b (solid line).  In
general, spectra measured on such positions showed fluctuations in
the degree of asymmetry, in the sharpness of the gap-edge
shoulders and in the possible existence of small zero-bias
anomalies (see below), but not in the general V-shaped structure
as expected for tunneling in the \emph{c} direction. This is due
to the finite tunneling-cone width in the STM experiment that
allows coupling to the electronic states in the \chem{CuO_2}
planes.  The general structure of this dI/dV-V curve is well
reproduced by a simulation based on the formalism for tunneling
into a \emph{d}-wave superconductor \cite{10, 11}, assuming
tunneling in the \emph{c} direction (dotted curve). In many cases
a small structure (far from being a well-developed ZBCP) appears
at around zero bias, as exhibited by the curve in fig.
\ref{Fig1}c, taken on top of a different crystallite. Similar
structures were found by other groups on \emph{c}-axis BSSCO
crystals and attributed to impurity-induced quasi-particle
excitation at the Fermi level \cite{20, 21, 23, 24}. Enhanced
gap-edge shoulders, as seen in the spectrum, have also been
observed previously \cite{3, 14, 21} and were related to a
van-Hove singularity in the electronic DOS \cite{14}.

The tunneling spectra obtained when positioning the tip above a
crystallite edge look considerably different when either the (100)
or the  (110) planes are probed. A nearly U-shaped gap structure
is obtained at position 2 (fig. \ref{Fig1}d), as well as in any
other position along the same facet, suggesting that the (100)
face is exposed at this edge, as confirmed by the relatively good
fit that takes into account such a contribution. The spectra
obtained in position 3 show a clear ZBCP inside a gap (fig.
\ref{Fig1}e). The ZBCP is attributed to ABS formed at the exposed
(110) facet. Typically, however, the tunneling spectra acquired on
larger (110) crystalline edges exhibited more pronounced ZBCP
structures, as depicted in fig. \ref{Fig1}f.  The fit to the curve
in fig. \ref{Fig1}e was obtained by taking contributions of (100),
(110) and (001) faces, consistent with the corner position (3 in
frame a), while that in fig. \ref{Fig1}f has only a (110)
component.

The main topographic features observed on films that were
deposited using a higher laser power were islands and voids of
unit-cell height on the (001) surface, as shown in image
\ref{Fig2}a. Image \ref{Fig2}b focuses on a corner between two
steps, each of a single unit-cell height, around which the spectra
presented in figures \ref{Fig2}c-\ref{Fig2}d were measured.
Generally, tunneling spectra taken far enough (a few nm) from a
step or a surface defect (such as holes of sub unit-cell depth
appearing in the image as black dots) exhibited a V-shaped
structure, such as curve 1 in fig. \ref{Fig2}c.  In the vicinity
of steps and "point-defects", ZBCPs of different magnitudes
embedded in relatively symmetric gap structures were observed.
This is exhibited by curve 2 (fig. \ref{Fig2}c), taken near a
(100) mono-step facet and showing a U-shaped gap, and by curve 3
(fig. \ref{Fig2}d), acquired on a (110) mono-step facet,
portraying a ZBCP within a V-shaped gap. The difference in the
apparent gap between curves 1 and 2 may be due either to local
variations in the oxygen content or to a life-time broadening
effect.  Such variations were observed also in STM measurements on
single crystals \cite{3}. To demonstrate the typical variance of
the spectra within the area shown in image \ref{Fig2}(b) we plot
in the insets of (c) and (d) additional characteristics acquired
at analogous positions to those presented in the main frames. It
is evident that the overall shapes are reproducible along a
specific topographic feature. To the best of our knowledge, this
is the first time that a clear correlation between tunneling
spectra and unit-cell step structure and orientation is
demonstrated. This indicates that even a single step of one
unit-cell height can alter considerably the \emph{measured} local
quasi-particle excitation, pointing out to the apparent
discrepancy between theory and experiment, discussed above.

We have also observed tunneling spectra that deviate from the
standard \emph{d}-wave behavior. For example, highly asymmetric
ZBCP structures were found on the same sample in similarly looking
areas (fig. \ref{Fig2}e). Such spectra were predicted to result
from scattering by atomic-scale impurities \cite{22, 24}.
Unfortunately, we can not yet specify any typical defect
structure, or any special topographic feature, which can be
directly related to the appearance of these highly asymmetric
tunneling spectra.

\begin{figure}
\onefigure [scale=0.75]{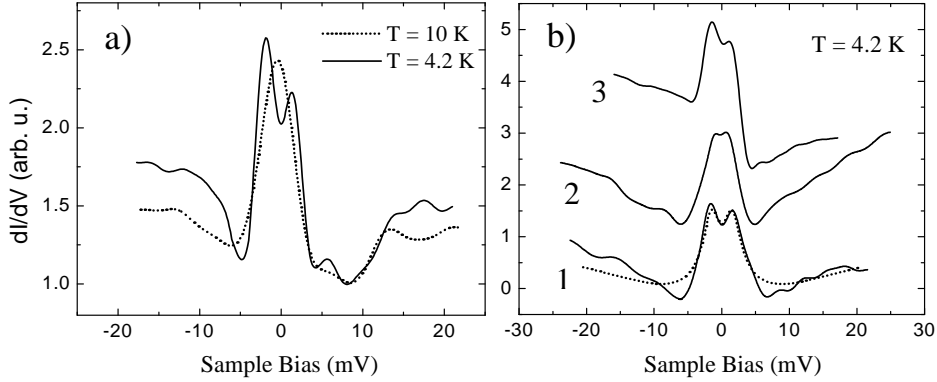} \caption{Splitting of the
ZBCP. (a) Two tunneling spectra measured at the same position, one
at $4.2 ~K$ (solid line) and the other at $10 ~K$ (dotted). The
clear split seen at $4.2 ~K$ is completely washed out at $10 ~K$,
signifying a transition into a broken time reversal symmetry
state. (b) Three tunneling spectra showing (1) the largest
splitting we have detected (3.2 meV peak to peak), (2) the
smallest splitting (1.6 meV) and (3) a pronounced asymmetric
curve.  The dotted line is a fit assuming a state of broken time
reversal symmetry having a $d_{x^{2}-y^{2}}+is$ order parameter,
with $\Delta_{d} = 30 ~meV$ and $\Delta_{s} = 0.04 \Delta_{d}$.}
\label{Fig3}
\end{figure}

We now turn to the issue of splitting of the ZBCP, a manifestation
of a state with broken time reversal symmetry, which we have
observed only in overdoped and optimally doped films. In fig.
\ref{Fig3}a we present two tunneling spectra obtained at the same
position on a nominally overdoped film, one taken at $4.2 ~K$,
exhibiting a clear splitting of the ZBCP ($\sim 2.6 ~meV$ peak to
peak), and the other at $10 ~K$, where the split is completely
washed out. (The lateral position of spectra acquisition was
determined at each temperature via the topographic images,
although slightly different tip-sample separations may have
yielded a small difference in the background.) This temperature
dependence indicates that the splitting is due to a broken time
reversal symmetry phase. The peak to peak separation we have found
in all of our measurements ranged between $1.6 ~meV$ (curve 2 in
fig. \ref{Fig3}b) to $3.2 ~meV$ (curve 1). The theoretical fit to
curve 1 was obtained using a $d_{x^{2}-y^{2}}+is$ order parameter
\cite{12}, where the magnitude of the subdominant \emph{s}
component, $\Delta_s$, is $4 \%$ of the dominant one, $\Delta_d$.
We note that we cannot clearly distinguish between the $is$
\cite{12} and the $id_{xy}$ \cite{13} scenarios for the
subdominant order parameter, although recent experiments seem to
support the latter case \cite{15, 25}.

Interestingly, the splitting into two well resolved peaks we
typically observed at $4.2 ~K$ was, in many cases, more pronounced
than the splitting into two weakly resolved peaks measured in
macroscopic planer junction at zero field \cite{4, 5}.  The reason
for this is that in the macroscopic tunneling junctions the
dI/dV-V curves display an average over large areas, whereas the
STM allows probing of specific locations where the splitting is
more pronounced.  Local probing enables also the observation of
highly asymmetric spectra that exhibit a state of broken time
reversal symmetry, such as curve 3.

The fact that splitting was not observed in underdoped samples
indicates that spontaneous time reversal symmetry breaking may
take place only beyond some critical doping level, as pointed out
by Deutscher \cite{15}.  This may be the reason for the
contradictory reports of various groups regarding the observation
of splitting in zero magnetic field, discussed above.  The spatial
variations in magnitude of the splitting seen in our nominally
overdoped and optimally doped samples may thus be due to
fluctuations in the \emph{local} hole doping.  The wide
superconducting transition exhibited by our samples after
completing the STM measurements, as compared to the freshly
prepared films, indeed indicates that the samples became
non-homogeneous electronically, probably due to inhomogeneous
oxygen distribution.  Clear evidence for spatial fluctuation in
the hole doping was previously observed by STM in other high-$T_c$
superconducting systems \cite{16, 17}.

In summary, tunneling spectroscopy measurements on ultra-thin
epitaxial YBCO films reveal a rich variety of \emph{d}-wave
spectral structures that show pronounced spatial variations with
nearly one-to-one correspondence to the surface morphology.  Even
a single unit-cell step is found to significantly affect the local
spectra.  The doping and position dependence of the ZBCP splitting
suggest that doping plays a major role in the transition to a
state with broken time reversal symmetry, which can occur only
beyond some doping level close to optimal doping.

\acknowledgments We are grateful to Guy Deutscher for useful
discussions. This research was supported in parts by the Israel
Science Foundation, the Heinrich Hertz Minerva center for high
temperature superconductivity and the Technion fund for the
promotion of research.


\begin{thebibliography}{0}
\bibitem{1}
  \Name{Van Harlingen D. J.}
  \REVIEW{Rev. Mod. Phys.}{67}{1995}{515}.
\bibitem{2}
  \Name{Orenstein J.\and Millis A. J.}
  \REVIEW{Science}{288}{2000}{468}.
\bibitem{3}
  \Name{Wei J. Y. T.,Yeh N. C.,Garrigus D. F. \and Starsik M.}
  \REVIEW{Phys. Rev. Lett.}{81}{1998}{2542}.
\bibitem{4}
  \Name{Covinton M., Aprili M., Paraonu E., Green L. H., Xu F., Zhu J. \and Mirkinet C.A.}
  \REVIEW{Phys. Rev. Lett.}{79}{1997}{277}.
\bibitem{5}
  \Name{Krupke R. \and Deutscher G.}
  \REVIEW{Phys. Rev. Lett.}{83}{1999}{4634}.
\bibitem{6}
  \Name{Iguchi I., Wang W., Yamazaki M., Tanaka Y. \and Kashiwaya S.}
  \REVIEW{Phys. Rev. B}{62}{2000}{R6131}.
\bibitem{7}
  \Name{Nesher O. \and Koren G.}
  \REVIEW{Phys. Rev. B}{60}{1999}{14893}.
\bibitem{8}
  \Name{Alff L., Beck A., Gross R., Marx A., Kleefisch S., Bauch Th., Sato H., Naito M. \and Koren G.}
  \REVIEW{Phys. Rev. B}{58}{1998}{11197}.
\bibitem{9}
   \Name{Wolf E. L.}
   \Book{Principles of Electron Tunneling Spectroscopy}
   \Publ{Oxford University press, New York}
   \Year{1985}.
\bibitem{10}
  \Name{Kashiwaya S., Tanaka Y., Koyanagi M. \and Kajimura K.}
  \REVIEW{Phys. Rev. B}{53}{1996}{2667}.
\bibitem{11}
  \Name{Tanaka Y.\and Kashiwaya S.}
  \REVIEW{Phys. Rev. Lett.}{74}{1995}{3451}.
\bibitem{12}
  \Name{Fogelstrom M., Rainer D., \and Sauls J. A.}
  \REVIEW{Phys. Rev. Lett.}{79}{1997}{281}.
\bibitem{13}
  \Name{Laughlin R. B.}
  \REVIEW{Phys. Rev. Lett.}{80}{1998}{5188}.
\bibitem{14}
  \Name{Wei J. Y. T., Tsui C. C., van Bentum P. J. M., Xiong Q., Chu C. W. \and Wu M. K.}
  \REVIEW{Phys. Rev. B}{57}{1998}{3650}.
\bibitem{15}
  \Name{Deutscher G., Dagan Y., Kohen A. \and Krupke R.}
  \REVIEW{Physica C}{341-348}{2000}{1629}.
\bibitem{16}
  \Name{Levi Y., Felner I., Asaf U.\and Millo O.}
  \REVIEW{Phys. Rev. B}{60}{1999}{R15059}.
\bibitem{17}
  \Name{Levi Y., Millo O., Sharoni A., Tsabba Y., Leitus G. \and Reich S.}
  \REVIEW{Europhys. Lett.}{51}{2000}{564}.
\bibitem{18}
  \Name{de Lozanne A. L.}
  \REVIEW{Supercond. Sci. Technol.}{12}{1999}{R43}.
\bibitem{19}
  \Name{Edwards H. L., Markert J. T. \and de Lozanne A. L.}
  \REVIEW{Phys. Rev. Lett.}{69}{1992}{2967}.
\bibitem{20}
  \Name{Hudson E. W., Pan S. H., Lang K. M., Gupta A. K., Ng K. W. \and Davis J. C.}
  \REVIEW{Physica B}{284-288}{2000}{969}.
\bibitem{21}
  \Name{Yazdani A., Howald C. M., Lutz C. P., Kapitulnik A. \and Eigler D.}
  \REVIEW{Phys. Rev. Lett.}{83}{1999}{176}.
\bibitem{new alff}
  \Name{Alff L.,Takashima H., Kashiwaya S., Terada N., Ihara H., Tanaka Y., Koyanagi
  \and Kajimura M.K.}
  \REVIEW{Phys. Rev. B}{55}{1997}{R14757}.
\bibitem{22}
  \Name{Zhu J.-X., Lee T. K., Ting C. S.\and Hu C. R.}
  \REVIEW{Phys. Rev. B}{61}{2000}{8667}.
\bibitem{23}
  \Name{Salkola M. I., Balatsky A. V. \and Scalapino D. J.}
  \REVIEW{Phys. Rev. Lett.}{77}{1996}{1841}.
\bibitem{24}
  \Name{Tanuma Y., Tanaka Y., Yamashiro M. \and Kashiwaya S.}
  \REVIEW{Phys. Rev. B}{57}{1998}{7997}.
\bibitem{25}
  \Name{Carmi R., Polturak E., Koren G. \and Auerbach A.}
  \REVIEW{Nature}{404}{2000}{853}.
\end{thebibliography}
\end{document}